# Hardware-algorithm collaborative computing with photonic spiking neuron chip based on integrated Fabry–Pérot laser with saturable absorber

Shuiying Xiang[1, 2*], Yuechun Shi[3*], Xingxing Guo[1], Yahui Zhang[1, 2], Hongji Wang[3], Dianzhuang Zheng[1], Ziwei Song[1], Yanan Han[1], Shuang Gao[1], Shihao Zhao[1], Biling Gu[1], Hailing Wang[4], Xiaojun Zhu[5], Lianping Hou[6], Xiangfei Chen[3], Wanhua Zheng[4], Xiaohua Ma[2] & Yue Hao[2]

**Photonic neuromorphic computing has emerged as a promising avenue toward building a low-latency and energy-efficient non-von-Neuman computing system. Photonic spiking neural network (PSNN) exploits brain-like spatiotemporal processing to realize high-performance neuromorphic computing. However, the nonlinear computation of PSNN remains a significant challenging. Here, we proposed and fabricated a photonic spiking neuron chip based on an integrated Fabry–Pérot laser with a saturable absorber (FP-SA) for the first time. The nonlinear neuron-like dynamics including temporal integration, threshold and spike generation, refractory period, and cascadability were experimentally demonstrated, which offers an indispensable fundamental building block to construct the PSNN hardware. Furthermore, we proposed time-multiplexed spike encoding to realize functional PSNN far beyond the hardware integration scale limit. PSNNs with single/cascaded photonic spiking neurons were experimentally demonstrated to realize hardware-algorithm collaborative computing, showing capability in performing classification tasks with supervised learning algorithm, which paves the way for multi-layer PSNN for solving complex tasks.**

[1] State Key Laboratory of Integrated Service Networks, Xidian University, Xi'an 710071, China. [2] State Key Discipline Laboratory of Wide Bandgap Semiconductor Technology, School of Microelectronics, Xidian University, Xi'an 710071, China. [3]Key Laboratory of Intelligent Optical Sensing and Manipulation, Ministry of Education, the National Laboratory of Solid State Microstructures, the College of Engineering and Applied Sciences, Institute of Optical Communication Engineering, Nanjing University, Nanjing 210023, China. [4]Laboratory of Solid-State Optoelectronics Information Technology, Institute of Semiconductors, Chinese Academy of Sciences, Beijing 100083, China. [5]School of Information Science and Technology, Nantong University, Nantong, Jiangsu, 226019, China. [6] James Watt School of Engineering, University of Glasgow, Glasgow, G12 8QQ, UK. Correspondence and requests for materials should be addressed to S.Y. X. and Y. C. S. (email: syxiang@xidian.edu.cn; shiyc@nju.edu.cn )

Nowadays, the conventional electronic processors based on the von Neumann's architecture are more and more difficult to maintain Moore's law. Inspired by the network architecture and principles of the human brain, neuromorphic computing has become one of the promising candidates to overcome the von-Neumann bottleneck in the post-Moore era [1]. Compared to the conventional continuous-value based artificial neural network (ANN) and convolutional neural network (CNN), the brain-inspired spiking neural network (SNN) provides a more biologically plausible way to implement neuromorphic computing and is believed to be much more powerful and lower power consumption due to the rich temporal information and event-driven manner [2-3]. In recent years, significant progress has been achieved in the electronic neuromorphic computing processors [4-6], such as SpiNNaker [7], Neurogrid [8], TrueNorth [9], Darwin [10], Loihi [11], and Tianjic [12], but they suffer from the limitations of computational speed and power consumption.

As an alternative, photonic technology offers huge hopes for next-generation neuromorphic processors, due to the fascinating advantages such as high speed, wide bandwidth, massive parallelism, and low power consumption [13-19]. The free-space [20-21] and the integrated photonics architectures [16-17] are two mainstream approaches for the photonic implementation of neural networks. Due to the potential of compact size and high reliability, the integrated photonics architecture is a promising approach and significant advancements have been achieved in recent years [22-30]. To use integrated photonics for neuromorphic computing, linear and nonlinear computation elements are indispensable fundamental building blocks and are of equal importance. The linear computation was successfully realized optically by Mach Zehnder interferometer (MZI) [22, 31-33], microring resonator (MRR) weight bank [23], phase-change material (PCM) integrated on waveguide [24-25], and semiconductor optical amplifier (SOA) [28]. However, among the majority of the existing photonic neural network chips, the nonlinear computation was implemented electronically rather than optically [22, 25, 28-33]. Thus, the optical implementation of nonlinearity remains one of the most challenging for optical neural networks.

In a photonic SNN (PSNN), the nonlinear computation is accomplished by the photonic spiking neuron. The spiking neuron is much more powerful than the continuous-value nonlinear activation due to the rich information represented with spikes based on the spatiotemporal processing mechanism. So far, the optical spiking neurons have been predicted numerically or demonstrated experimentally based on discrete devices [34-39] and integrated schemes [40-47]. But these approaches also face some limitations. For instance, the PCM-based spiking neuron lacks temporal integration function, which is crucial to the spike processing [24, 43]. The output power of micropillar laser neuron is relatively low, which may require additional amplification to compensate for the loss when applied to multi-layer or deep PSNN [44-45]. The integrated distributed feedback laser neuron requires optic-electronic conversion, which may cause increased system complexity and power consumption [46]. To pave the way for the practical application of PSNN, it is still highly desirable to explore a novel photonic spiking neuron chip that can take full advantage of the temporal encoding feature and can be easily applied to multi-layer or deep PSNN for solving complex tasks.

Here, we proposed and fabricated a photonic spiking neuron chip based on an integrated Fabry–Pérot laser with an intracavity saturable absorber (FP-SA) for the first time, which could form an integral part of hardware PSNN. Besides, to avoid the currently available photonics integration scale limit, we proposed time-multiplexed spike encoding to realize spatiotemporal encoding with a single photonic spiking neuron, which enabled the implementation of large-scale PSNN far beyond the hardware integration scale limit. Furthermore, the hardware-algorithm collaborative computing based on PSNN consisting of single and two cascaded photonic spiking neurons were experimentally demonstrated to perform the pattern classification task with a modified supervised learning algorithm, which represents a major step forward in the practical application of PSNN. Our experimental demonstrations present a promising avenue toward hardware-software co-design and optimization of large-scale multi-layer PSNN for solving complex tasks, and will be promising for wide applications such as in machine learning, artificial intelligence, data centers, edge computation, and autonomous driving.

## Results

**Operation principle of PSNN for pattern recognition with time-multiplexed spatial-temporal spike encoding mechanism.** Here, we proposed the time-multiplexed spike encoding with a single photonic spiking neuron to realize the brain-like spatial-temporal encoding. The classification task is to distinguish four simple number patterns (i.e., 1, 2, 3, and 4), with a 5×4 pixel matrix, as presented in Fig. 1 (a). As presented in Fig. 1 (b), we consider a small PSNN with four pre-synaptic (PRE) neurons and four post-synaptic (POST) neurons. Each column in the matrix of 5 pixels is encoded with a PRE neuron. The spike encoding for each input pattern can be expressed as $t_e(m) = I(m) \times [x+y+5]$ ns, in which $x$ and $y$ are the subscript index of elements in pixel matrix, respectively. $I(m)$ represents the $m$-th column pixel intensity encoded by the PRE-$m$ neuron, and the value is 0 or 1 corresponding to the pixel of white or black. For example, for input pattern "2", pixel intensity of each columns are $I(1)$=[1,0,1,1,1], $I(2)$=[1,0,1,0,1], $I(3)$=[1,0,1,0,1] and $I(4)$=[1,1,1,0,1], respectively. And the corresponding encoded spiking timings are $t_e(1)$=[7, 9, 10, 11] ns, $t_e(2)$=[8, 10, 12] ns, $t_e(3)$=[9, 11, 13] ns, $t_e(4)$=[10, 11, 12, 14] ns. All the PRE spikes are weighted and then propagated to four POSTs. The weight matrix consists of $\omega_{ij}$ which represents the synaptic weight between the $j$-th PRE and the $i$-th POST. Note, the weights are trained with a supervised algorithm in a computer, and only the inference process is accomplished in the hardware implementation in the present work.

The spatiotemporal pattern of target response is shown in Fig. 1 (c), and is defined as target = [1, 0, 0, 0; 0, 1, 0, 0; 0, 0, 1, 0; 0, 0, 0, 1]. Here, one row represents one POST, and one column denotes the target response of four POSTs corresponding to a specific input pattern. For simplicity, we use 1 and 0 to indicate that the POST fires and doesn't fire a spike, respectively. For instance, for pattern "1", the target is that POST1 fires a spike, and the rest three POSTs do not fire a spike. Thus, the first column [1, 0, 0, 0] represents the target response of pattern "1". For pattern "2", the target is that only

POST2 fires a spike, and the target response corresponds to the second column [0, 1, 0, 0]. Similarly, the target responses of pattern "3" and "4" correspond to the third column [0, 0, 1, 0] and the fourth column [0, 0, 0, 1], respectively.

In our proposed scheme, we used only one POST neuron to achieve the same functions of four POSTs of the entire layer of PSNN by the time-multiplexed spike encoding mechanism. As presented in Fig.1 (d), one POST neuron has four response time windows corresponding to four POSTs to respond to a specific input pattern. For example, as shown in Fig. 1 (e), for pattern "1", the POST only fires a spike in the first time window and does not fire in the rest of three time windows. Thus, the response of the POST can be represented as [1, 0, 0, 0]. For pattern "2", "3" and "4", the POST only fires a spike in the second, third and fourth time window, and the response of the POST can be denoted as [0, 1, 0, 0], [0, 0, 1, 0] and [0, 0, 0, 1], respectively.

The nonlinear computation functions of the POST are accomplished by the proposed photonic spiking neuron, which mainly include temporal integration, threshold, spike generation, and refractory period, as presented in Fig.1 (f). In our work, the photonic spiking neuron chip based on the integrated FP-SA was designed, optimized, and fabricated. The integrated FP-SA laser chip includes two electrical isolation sections, i.e., a gain region and a saturable absorber (SA) region. Fig. 1(g) shows the microscopic image of the fabricated laser. The total length of the chip is $L_{cavity}$=1500 μm, the width of the laser chip is 300 μm, and the ridge waveguide width is 2.5 μm. We designed and fabricated the devices with four different SA lengths, $L_{SA}$= 25 μm, 30 μm, 75 μm, and 90 μm. The SA section side facet and gain section side facet are coated with reflections of 95% and 30% respectively. For clarity, we denote this photonic spiking neuron chip based on the two-section integrated semiconductor laser as FP-SA in the following. For the FP-SA with different SA lengths, we packaged several devices to demonstrate uniformity. The material used in this work is a p-i-n diode structure grown based on the AlGaInAs/InP material system with an epitaxial layer structure described in[48，49].

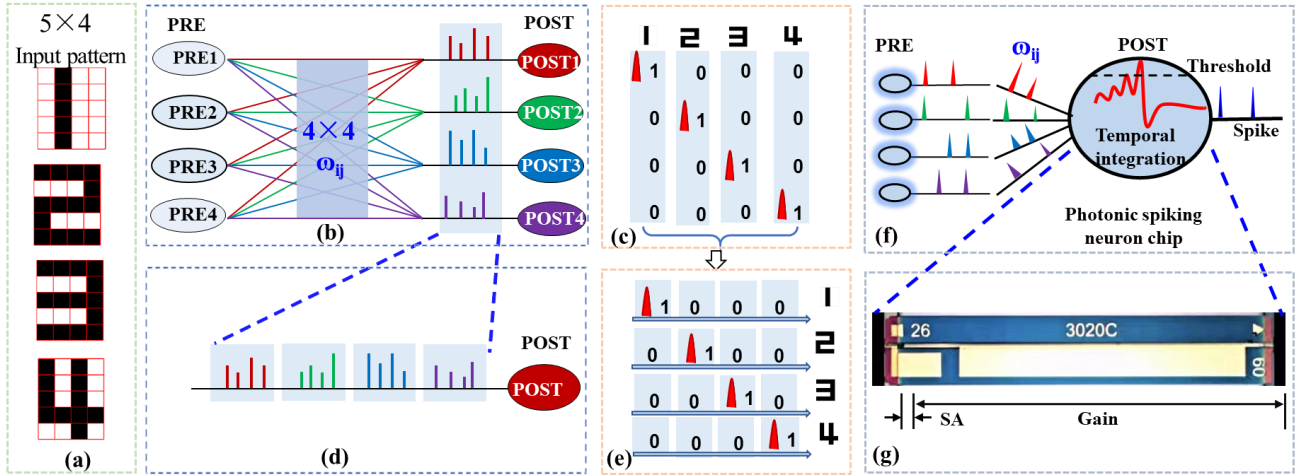

Fig. 1. The operation principle of PSNN for patter recognition. (a) The input patterns with 5×4 pixel matrix, (b) PSNN with 4 input nodes and 4 output nodes, (c) the spatial-temporal target output response for recognition of different input patterns; (d) the proposed time-multiplexed spike encoding to realize the spatial-temporal spike encoding with a single FP-SA neuron, (e) the target output response for different input patterns, (f) the nonlinear computation mechanism of a photonic spiking neuron chip, (g) The chip layout of the integrated FP-SA chip. PRE: pre-synaptic spiking neuron; POST: post-synaptic spiking neuron. $\omega_{ij}$ is the synaptic weight between the $j$-th PRE and the $i$-th POST, which is trained with supervised learning algorithm.

**Experimental setup for photonic spiking neuron.** The experimental setup of an FP-SA neuron for emulating the neuron-like dynamics is presented in Fig. 2 (a). An arbitrary waveform generator (AWG, Tektronix AWG70001A) produced the electronic stimulus. A tunable laser (TL, AQ2200-136 TLS module) provided the optical carrier. The electro-optical conversion was realized with an MZM. Here, the half-wave voltage of the MZM was 1.9 V. The modulated optical stimulus was then injected into the facet of the gain section of the FP-SA via an optical circulator. The gain region was forward biased with a laser diode controller (LDC: ILX Lightwave LDC3724B) that provided low-noise bias current and precise temperature control. The saturable absorber was reversed biased with the voltage source (VS). The variable optical attenuator (VOA) was used to adjust the power of the injected signal from MZM and EDFA. The polarization controllers (PC1 and PC2) were employed to match the polarization state. The output of the FP-SA neuron was analyzed by an optical spectrum analyzer (OSA, Advantest Q8384), and meanwhile, was detected by a photodetector (PD, Agilent/HP 11982A) and then recorded by an oscilloscope (OSC, Keysight DSOV334A) and an RF spectrum analyzer (Rohde & Schwarz FSW85).

In the experiment, when the temperature was fixed at 25°C, the power-current (PI) curves for free-running FP-SA lasers are presented in Figs. 2(b1) and (b2) under different cases of inverse bias of the SA region. The PI curves for four FP-SA lasers with different $L_{SA}$ are similar. Here, only the PI curves for FP-SA with $L_{SA}$= 25 μm and $L_{SA}$ =30 μm are presented. For both devices, it can be seen that the threshold of gain current is $I_G$=30 mA for reverse voltage $V_{SA}$ =0

V. An increase of the SA reverse voltage raises the threshold current and reduces the slope efficiency, as a result of an increase in interband and exciton absorption [48].

**Self-pulsation regimes of FP-SA.** Note, for the free-running FP-SA, when the gain current $I_G$ and reverse voltage $V_{SA}$ are sufficiently large, two different stable pulse operation regimes can be observed, i.e., pure mode-locking and self-pulsation regions. For example, for the FP-SA with $L_{SA}$ =75 μm, when $I_G$=95 mA and $V_{SA}$ = -3.05 V, the optical spectrum and RF spectrum are shown in Figs. 2 (c1) and (c2). Multiple longitudinal modes with clear and deep modulation (maximum modulation > 45 dB) can be seen, and the mode spacing is about 0.24 nm. This corresponds to a frequency spacing of 28.9 GHz, which is equal to $c/(2n_g L_{cavity})$. Here, $n_g$ =3.46 is the group refractive index of the ridge waveguide of the FP-SA chip, and $c$ is the speed of light in a vacuum. The RF spectrum shows that the FP-SA emits pulses with a fundamental frequency of 28.9 GHz and their high order harmonic components. When the reverse voltage is fixed at $V_{SA}$ = -3.05 V, the FP-SA emits around 28.9 GHz pulse trains for the gain current range from 89.2 to 98 mA. These spectra properties indicate that the FP-SA operates at the pure mode-locking regime with 100% modulation. Besides, when we further increase the $I_G$ and $V_{SA}$, the FP-SA enters the self-pulsation regime. For $V_{SA}$= -4.49 V, the frequency varies from 1.49 GHz to 1.95 GHz when $I_G$ is increased from 100 to 120 mA, which is due to the Q-switching mechanism. The time series, optical spectrum and RF spectrum are presented in Figs. 2(d1), (d2) and (d3) for $I_G$ =100 mA. Compared to the pure mode-locking, the optical spectrum widens while its modulation depth decreases. In our experiments, we find that the two different pulse regimes are more critical to the $V_{SA}$ but can be maintained for a relatively wide range of $I_G$.

For all the considered FP-SA lasers, two pulse regimes can all be observed, but with slightly different $I_G$ and $V_{SA}$ ranges. Note, to emulate the neuron-like dynamics with the FP-SA, the bias current and reverse voltage should be set below the pulse threshold. In such a case, the FP-SA operates at the excitability regime, and small external input light perturbations can excite the FP-SA to produce a single pulse [50].

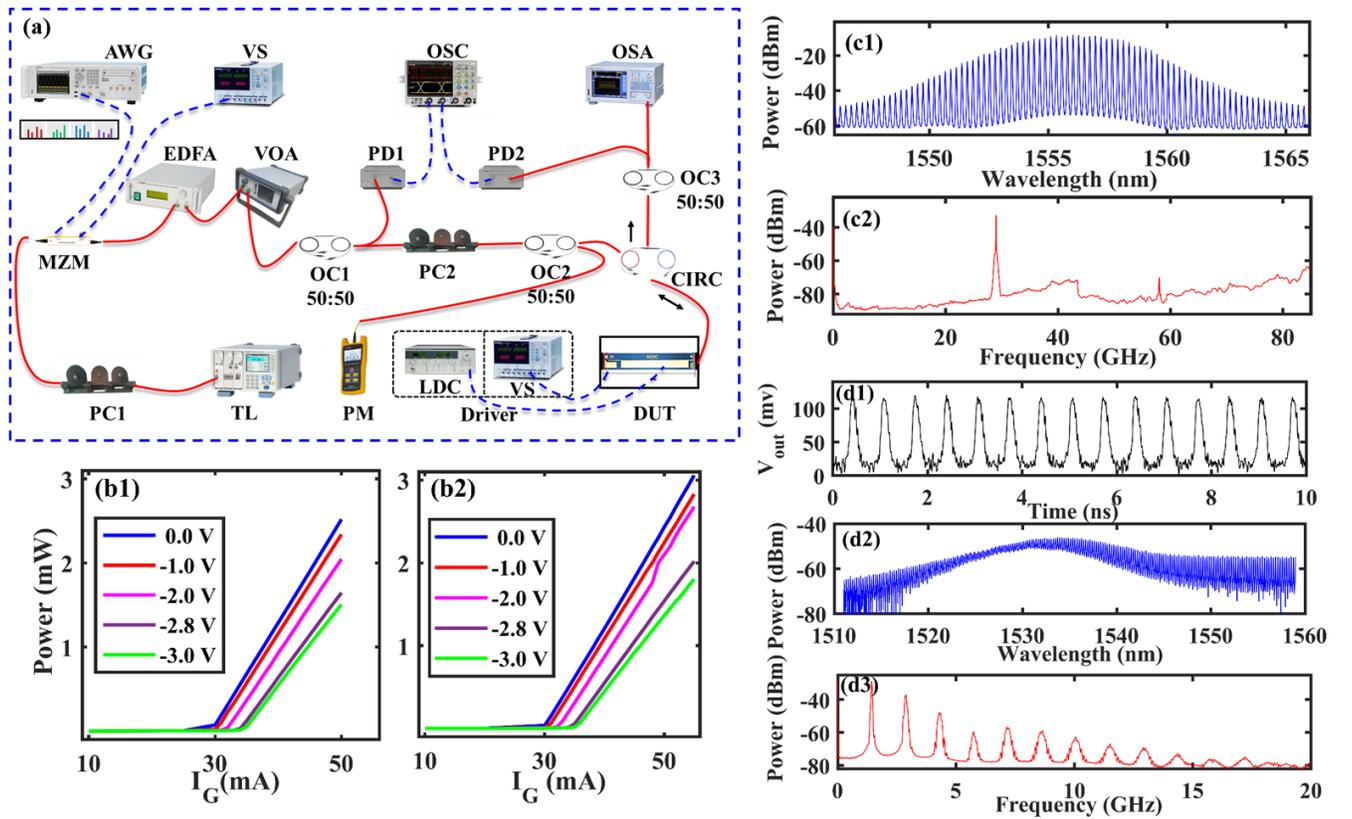

Fig. 2. Experimental setup, PI curves and the self-pulsation output characteristics of FP-SA neuron. **(a)** The experimental setup for testing the FP-SA neuron. AWG: arbitrary waveform generator; VS: voltage source; TL: tunable laser; OI: optical isolator; VOA: variable optical attenuator; PC1 and PC2: polarization controllers; MZM: Mach-Zehnder modulator; OC1 and OC2: optical couplers; CIRC: circulator; PD1 and PD2: photodetectors; PM: power meter; OSC: oscilloscope; OSA: optical spectrum analyzer; LDC: laser diode controller; DUT: device under test. PI curve of FP-SA with (b1) $L_{SA}$= 25 μm and (b2) with $L_{SA}$=30 μm. (c1) Optical spectra and (c2) RF spectrum of free-running FP-SA with $L_{SA}$= 75 μm, $I_G$=95 mA, $V_{SA}$ = -3.05 V. (d1) Time series, (d2) optical spectra, and (d3) RF spectrum of free-running FP-SA with $I_G$=100 mA, $V_{SA}$ = -4.49 V.

**Nonlinear neuron-like dynamics of photonic spiking neuron based on the FP-SA.** We designed the perturbation signal generated by the AWG to demonstrate the complex nonlinear neuron-like dynamics of the photonic spiking neuron. Here, we only present the results for the FP-SA with $L_{SA}$= 30 μm. As shown in Figs.3 (a1) and (a2), the perturbation signal includes three pulses with different power, only the first injected pulse triggers the FP-SA neuron to generate a neuron-like spike. While the responses of the FP-SA neuron to the second and the third perturbation pulses are negligible, which indicates the excitability threshold. Besides, we also designed three types of pulse stimuli to demonstrate the temporal integration effect, as shown in Figs. 3(b1) and (b2). The first pulse with high input power elicits a spike generation of the FP-SA neuron, the second pulse burst with three closely spaced weak pulses with the inter-spike interval (ISI) of 500ps, also triggers a spike generation, while the third single sub-threshold perturbation pulse does not elicit the response spike. That is to say, even a single sub-threshold pulse cannot reach the spike threshold, the three closely-spaced sub-threshold pulses are temporally integrated and thus exceed the threshold, which demonstrates the temporal integration property of the FP-SA neuron. As presented in Figs.3 (c1) and (c2), we designed and generated a burst of pulse with relatively a large ISI (the ISI is 2ns). We can find that the first perturbation pulse can elicit a response spike, indicating that the single perturbation pulse power exceeds the threshold. However, the second perturbation pulse cannot trigger another response spike because for the gain section to fully recover its gain, it takes time in the order of several ns [51]. The bursts of five pulses only trigger three response spikes, which indicates the refractory period [44]. Obviously, the response speed is much faster than the biological counterpart. We further present the experimental colour-coded temporal maps plotting superimposed time series of the responses corresponding to 100 consecutive arriving stimuli events, as in Figs.3(d1), (d2), and (d3). For the three cases, the same spiking response is obtained for every single one of the 100 incoming stimuli. Hence, reproducible spiking responses can be obtained in our fabricated FP-SA neuron. The optical spectrum for the optically injected FP-SA that operates as a photonic spiking neuron is further presented in Fig. 3(e). Here, the wavelength of the external optical stimulus is 1561.48 nm, which is far away from the peak wavelength of the FP-SA, i.e., 1554.38 nm. Namely, incoherent optical injection is desired to realize neuron-like dynamics.

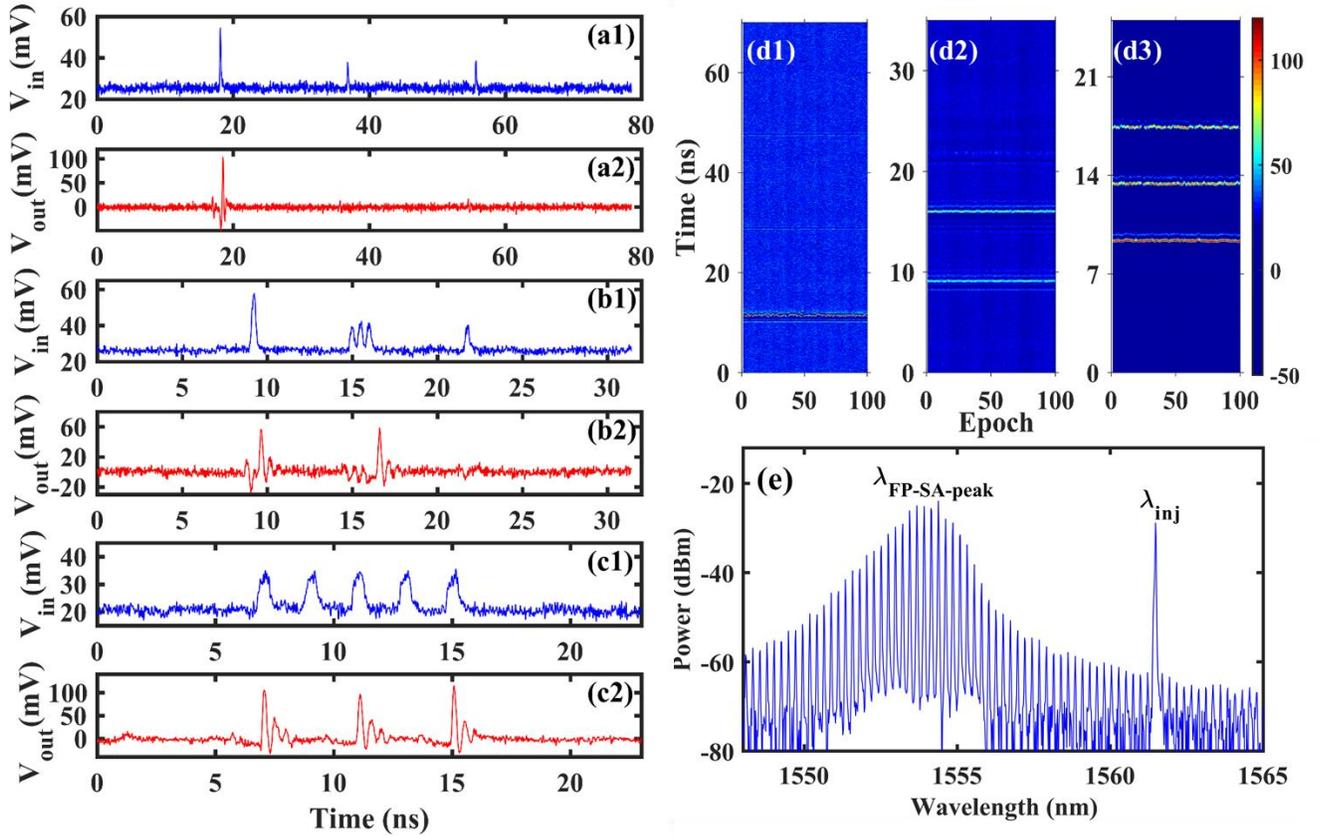

Fig. 3. Experimental demonstration of neuron-like dynamics and the corresponding optical spectrum. (a) Threshold, (b) temporal integration, (c) refractory period. (a1, b1, c1) represent the stimulus signal of the laser neuron, (a2, b2, c2) denote the response of the FP-SA neuron. The FP-SA with $L_{SA}$=30 μm is considered. Temporal maps plotting the response of laser neuron to the arrival of 100 consecutive external stimuli, (d1) corresponds to stimulus in (a1), (d2) corresponds to stimulus in (b1), (d3) corresponds to stimulus in (c1). (e) The optical spectra for the FP-SA operates as a photonic spiking neuron.

**Training process based on modified supervised learning algorithm.** The model and modified supervised training algorithm are presentd in section Method. The training process and the simulation results are shown in Fig. 4. For each pattern, the training convergence is achieved after several epochs. For pattern "1", only POST 1 emits a spike at 10ns, while the rest three POSTS emit no spike, and the timing is the maximum of the training window. Similarly, for patterns "2", "3", and "4", the training convergence agrees well with the defined target. The weights after training convergence are presented in Fig. 4 (b). After the training convergence, the spike encoding output of each PRE is multiplied by the trained weight matrix. To intuitively present the insight into the weight process, the weighted signals that are injected into each POST for patterns "3" and "4" are shown in Figs. 4 (c) and (d). The corresponding responses of each POST for patterns "3" and "4" are shown in Figs. 4 (e) and (f). Obviously, only POST 3 emits a spike for pattern "3". For pattern "4", only POST 4 generates a spike. Note that, to realize the time-multiplexed spike encoding, as shown in Fig.1 (d), the weighted additions are combined to obtain a single stimulus signal that is imported to the hardware photonic spiking neuron. As this work focuses on the hardware implementation of a photonic spiking neuron, the combined weighted signal is directly mapped to the output of AWG.

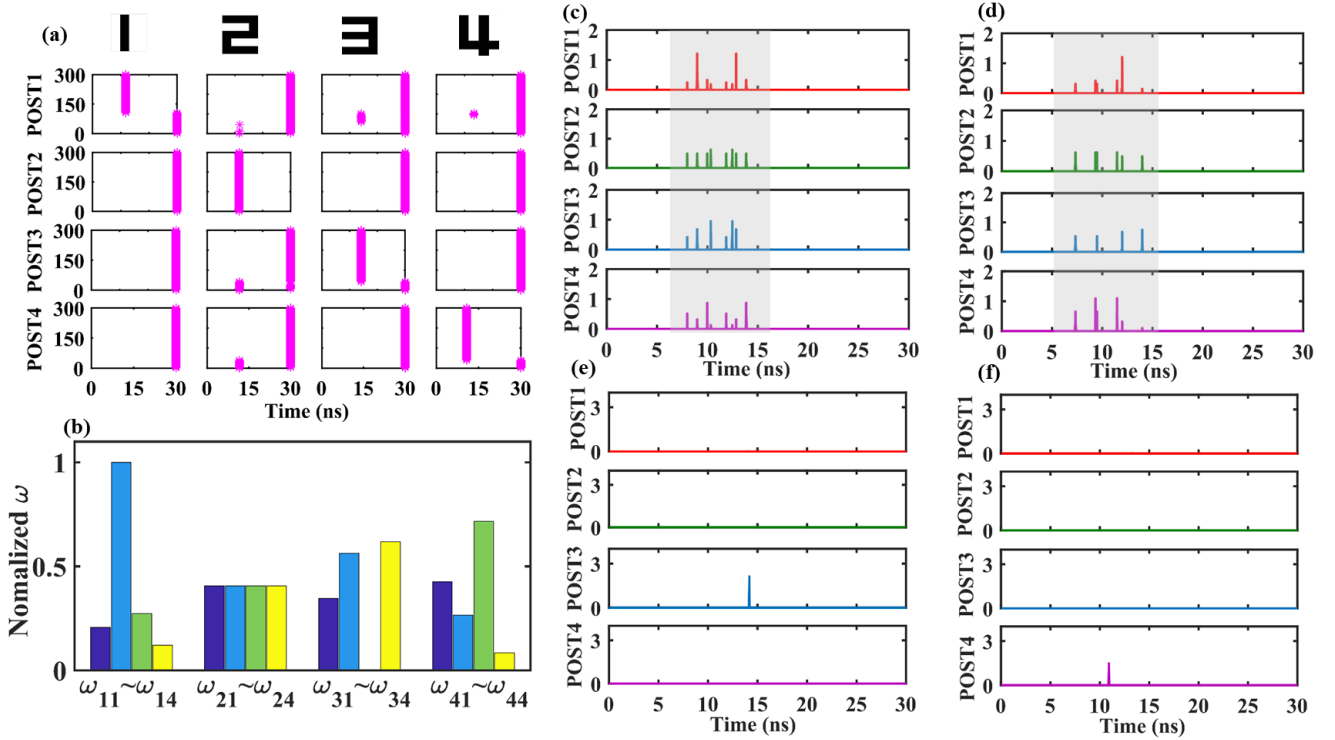

Fig. 4. The training process and simulation results based on PSNN. (a) the training process for each pattern, (b) weight after training convergence, (c) and (d) represent the inputs of each POST for the pattern "3" and "4", respectively, (e) and (f) represent the response of each POST corresponding to (c) and (d). $\omega_{ij}$ denotes the synaptic connection weight between $j$-th PRE and $i$-th POST.

**Hardware implementation of PSNN based on single FP-SA.** After training convergence, the weighted signal for each POST is mapped to a time-multiplexed signal that can be generated by the AWG. The electronic output of AWG is modulated by the MZM and injected into the fabricated FP-SA to realize a photonic spiking neuron. When the input pattern is "1", the time-multiplexed input for the POST is presented in Fig. 5 (a1). As seen in Fig. 5 (a2), only one spike appears in the first time window, while the FP-SA laser neuron responds with sub-threshold oscillation during the rest time windows, which agrees well with the target output as denoted in Fig. 1 (e). Hence, the pattern '1' is successfully recognized by the spatiotemporal dynamics of the single photonic spiking neuron. When the input pattern is '2', the time-multiplexed weighted addition signal presented in Fig. 5 (b1) is quite different from that of pattern '1', and the FP-SA laser neuron responds with a high-intensity spike during the second time window as shown in Fig. 5 (b2). Similarly, as shown in Figs. 5 (c1) - (d2), when the input pattern is "3" ("4"), the FP-SA laser neuron responds with a high-intensity spike during the third (fourth) time window. Thus, the inference process of the classification task is successfully demonstrated in the hardware. Here it is worth pointing out that the hardware-algorithm collaborative computing based on PSNN with the time-multiplexed spike encoding is realized for the first time.

As displayed in Fig. 6, the colour-coded temporal maps plotting superimposed time series of the responses correspond to 500 consecutive arriving stimuli events for the four patterns. It can be seen that the same spiking response is obtained for each pattern. Note that, even the electronic noise is inevitable in the AWG, PD, and OSC, as well as the environment variation and the FP-SA laser neuron's noise, the inference process is still robust. Hence, reproducible pattern classification results can be achieved with the fabricated photonic spiking neuron based on the FP-SA.

**Hardware implementation of multi-layer PSNN with two cascaded photonic spiking neurons.** We further cascade two hardware photonic spiking neurons to study the cascadability property and multi-layer PSNN. As presented in Fig. 7 (a), FP-SA1 represents the first photonic spiking neuron, and FP-SA2 denotes the second photonic spiking neuron. The output of FP-SA1 is optically injected into FP-SA2, and the injected optical power can be adjusted by a VOA between them. As shown in Figs. 7 (b) and (c), the neuron-like threshold and temporal integration property can be achieved in both FP-SAs, which demonstrate the cascadability of a biological neuron is successfully emulated.

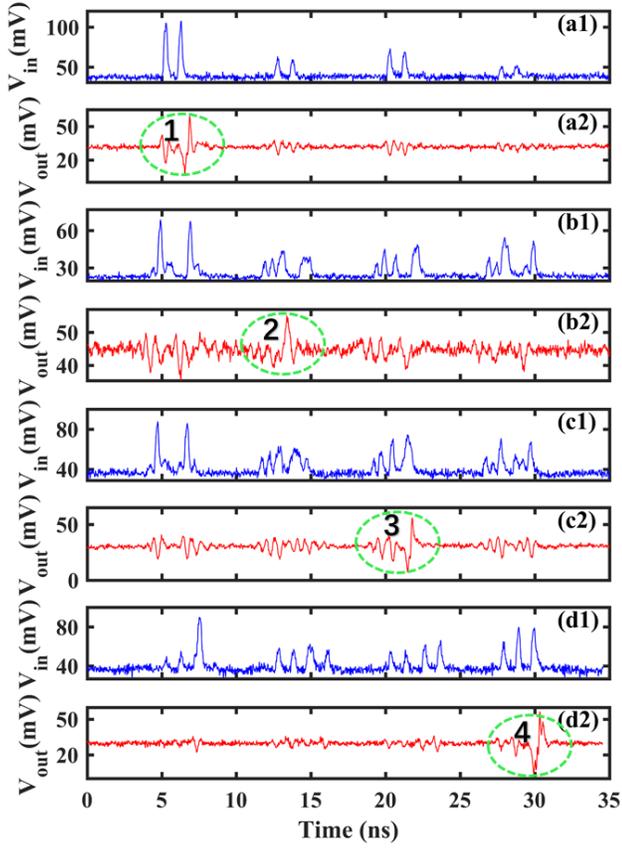

Fig. 5. The response of the fabricated photonic spiking neuron for different input patterns. (a1) represents the time-multiplexed input of photonic spiking neuron for pattern "1" and (a2) is the corresponding response, (b1) and (b2) correspond to pattern "2", (c1) and (c2) correspond to pattern "3", (d1) and (d2) correspond to pattern "4". The FP-SA with $L_{SA}$=30 μm is considered.

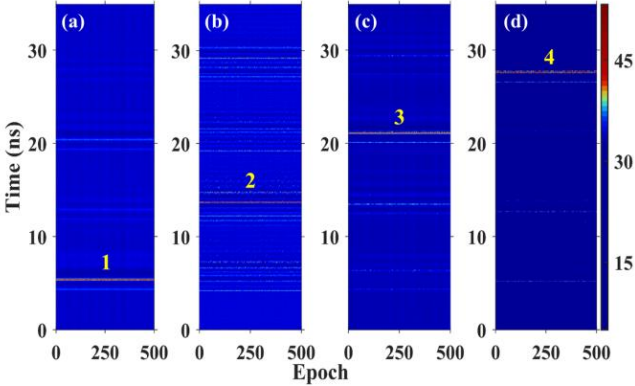

Fig. 6. Temporal maps plotting the response of photonic spiking laser neuron to the arrival of 500 consecutive external stimuli. (a) - (d) corresponds to pattern "1", "2", "3", and "4", respectively.

With the cascaded configuration, we also consider the pattern recognition tasks. The inference performances of "XDU" and "NJU" are presented in Fig. 8. The input patterns are represented by a 5×5 pixel matrix. The PSNN with 5 pre-synaptic FP-SA neurons and 3 post-synaptic FP-SA neurons is employed. The input spike encoding method and training algorithm are the same as those for the 4×4 PSNN. Here, the target response of POST1, POST2 and POST3 is [1, 0, 0], [0, 1, 0] and [0, 0, 1], respectively. After training convergence, the weight matrix is multiplied by the spike output of 5 PREs, and then the time-multiplexed spike encoding signal is optically injected into the FP-SA1, and the output of FP-SA1 is optically injected into FP-SA2. For both tasks, three input patterns can be successfully recognized by two cascaded photonic spiking neurons based on FP-SA with desired spike emission window.

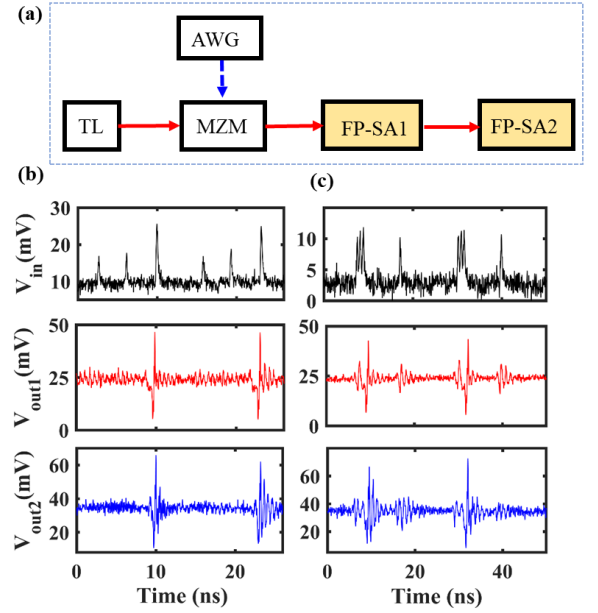

Fig. 7. The cascaded photonic spiking neurons and the cascadability. (a) Schematic diagram of cascaded two hardware photonic spiking neurons, FP-SA1 with $L_{SA}$=75 μm, FP-SA2 with $L_{SA}$ =90 μm. The cascadability of the threshold (b) and temporal integration (c). $V_{in}$ means the input of FP-SA1, $V_{out1}$ denotes the response of FP-SA1, $V_{out2}$ represents the response of FP-SA2. For FP-SA1, $I_G$=60.5 mA and $V_{SA}$= -2.34 V, T=23.5 ℃; for FP-SA2, $I_G$=58.4 mA and $V_{SA}$= -2.56 V, T=24.7 ℃.

Interestingly, we find that, better recognition performance can be achieved by the FP-SA2. On the one hand, from the physical perspective, such better neural computation performance may take advantage of the combined saturable absorption effects of two FP-SAs. On the other hand, from the neural network perspective, this may be because the cascaded configuration exhibits a multilayer PSNN as each FP-SA represents an entire layer of spiking neurons with the time-multiplexed spike encoding.

**Discussions**

In this work, we fabricated a novel photonic spiking neuron chip based on an integrated FP-SA for the first time, which has a simple structure and can be easily integrated on large scale with a commercially mature semiconductor process. The reproducible complex nonlinear neuron-like dynamics such as temporal integration, threshold and spike generation, and refractory period were demonstrated experimentally and are much faster than their biological and electronic counterparts. The proposed approach offers a novel indispensable fundamental building block to realize the PSNN hardware, and plays a key foundation for large-scale integrated PSNN chips. Furthermore, we proposed time-multiplexed spike encoding to mimic the brain-like spatiotemporal processing, and realized the hardware implementation of the inference process for pattern classification tasks with a single photonic spiking neuron based on a modified supervised learning algorithm. Note that, the time-multiplexed spatiotemporal spike encoding enabled the implementation of large-scale functional PSNN far beyond the hardware integration scale limit. A multi-layer PSNN with two cascaded photonic spiking neurons was also realized experimentally, both the cascadability performances and the pattern recognition inference tasks were demonstrated successfully. The first experimental demonstration of hardware-algorithm collaborative computing based on photonic spiking neuron represents a major step towards prompting the practical application of integrated PSNN chip, and proves the potential to build a large-scale multi-layer PSNN chip for solving complex tasks.

In the experiments, we find that temperature plays a key role in neuro-inspired computing as it determines the frequency detuning between the injected light and the longitudinal mode of the FP-SA laser neuron. Hence, temperature management should be emphasized when considering the large-scale photonic spiking neuron array. To further improve the inference performance, the hardware-software co-design and optimization of the PSNN are highly desirable in the future iteration.

As a further attempt, it is highly desirable to integrate the III-V laser neuron chip with the silicon-photonics-based weight devices such as MRR or MZI network, or the InP-based weight devices such as SOA. There are lots of experimental demonstrations on the weight matrix multiplication in the past few years [16]. It is suggested that a hybrid III-V/ Silicon integration may be a promising solution for the implementation of the entire on-chip PSNN [52-54]. Besides, the fast time-varying weight matrix is also deserved further innovation to take full advantage of the time-multiplexed spatiotemporal spike encoding mechanism of photonic spiking neuron.

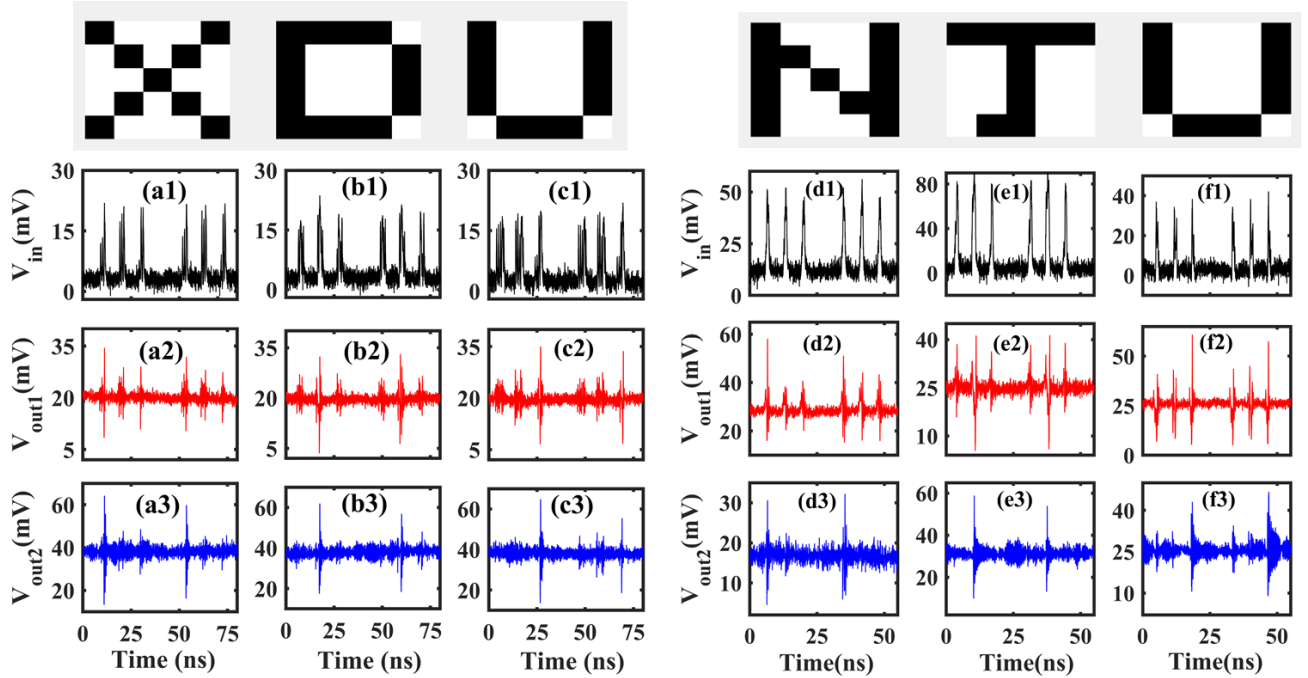

Fig. 8. Pattern recognition of patterns with cascaded photonic spiking neurons. (a-c) "X", "D", "U" and (d-f) "N", "J", "U".

**Method**

**Model and algorithm.** To design the supervised learning algorithm and realize the hardware-algorithm co-design, we model the FP-SA with rate equations based on the Yamada model [26-27]:

$$\frac{dS_{i,o}}{dt} = \Gamma_a g_a (n_a - n_{0a}) S_{i,o} + \Gamma_s g_s (n_s - n_{0s}) S_{i,o} - \frac{S_{i,o}}{\tau_{ph}} + \beta B_r n_a^2 \quad (1)$$

$$\frac{dn_a}{dt} = -\Gamma_a g_a (n_a - n_{0a})(S - \Phi_{pre,i} - \Phi_{post,o}) - \frac{n_a}{\tau_a} + \frac{I_a}{eV_a} \quad (2)$$

$$\frac{dn_s}{dt} = -\Gamma_s g_s (n_s - n_{0s}) S_{i,o} - \frac{n_s}{\tau_s} + \frac{I_s}{eV_s} \quad (3)$$

The parameters used in simulations are the same as those in Ref. [27].

During the training process of the PSNN, a modified tempotron-like ReSuMe supervised learning algorithm is used [55-56], and can be described as follows,

$$\Delta\omega_{ij} = \begin{cases} \sum_{t_i \leq t_{\max}} K(t_{\max} - t_i), & \text{if } n^d = 1, n^o = 0 \\ -\sum_{t_i \leq t_{out}} K(t_{out} - t_i), & \text{if } n^d = 0, n^o = 1 \\ 0, & \text{if } n^d = n^o \end{cases} \quad (4)$$

$$\omega_{ij}(x+1) = \omega_{ij}(x) + \omega_f \times \Delta\omega_{ij} \quad (5)$$

The $K$ function can be expressed as:

$$K(t) = V_0 \cdot \left( \exp\left(\frac{-t}{\tau_m}\right) - \exp\left(\frac{-t}{\tau_s}\right) \right) \quad (6)$$

In which, $V_0 = 2.1165$, $\tau_m = 1ns$, $\tau_s = \tau_m / 4$. This learning window only considers spikes $t_i \leq t$. $\omega_{ij}(x)$ and $\omega_{ij}(x+1)$ are the synapse weights at $x$-th and $(x+1)$-th learning epochs, respectively. $\omega_f$ is the learning rate and is set as $0.4 \times 10^8$. It denotes the maximum change in synaptic efficacies. $\Delta\omega_{ij}$

represents the weight update amount. $n^d$ and $n^o$ are the number of spikes from the desired and the actual output spike trains, respectively. The algorithm updates its weights whenever the neuron fails to respond as the same desired state as the teacher. When $n^d = 1$ is presented to the POST, it should fire a spike. However, once the actual response of POST is no spike ($n^{out} = 0$), the synaptic weight should be strengthened. On the other hand, when $n^d = 0$ is presented to the POST, it should keep silent. If the actual response of POST fires a spike ($n^{out} = 1$), the synaptic weight should be depressed. Here, the shape of the learning window follows kernel $K$ and the changing amount of the weight depends on the time difference between $t_i$, $t_{out}$ and $t_{max}$. In which $t_i$ is the pre-synaptic spike time, $t_{out}$ is the actual output spike time, and $t_{max}$ denotes the time at which the neuron reaches its maximum output power value in the time domain.

**Data availability**. The data that support the findings of this study are available from the corresponding author upon request.

## Acknowledgements

This work was supported by the National Key Research and Development Program of China (2021YFB2801900, 2021YFB2801901, 2021YFB2801902, 2021YFB2801904); National Natural Science Foundation of China (No. 61974177, No.61674119); National Outstanding Youth Science Fund Project of National Natural Science Foundation of China (62022062); The Fundamental Research Funds for the Central Universities (JB210114). We would like to thank Dr. Jianji Dong, Dr. Qiang Li, and Dr. Cuicui Lu for their helpful discussions.

## Author contributions

S. Y. X. and Y. C. S. designed the experiments. Y. C. S., H. J. W., H. L. W., L. P. H., X. F. C., and W. H. Z. designed and fabricated the devices. X. X. G., Y. H. Z., D. Z. Z., B. L. G. and S. H. Z. performed experimental measurements. Z.W. S. and Y. N. H. performed the simulations. S. Y. X., Y. C. S., X. J. Z., and L. P. H. prepared the manuscript. S. Y. X., X. H. M., and Y. H. directed all the research. All authors analyzed the results and implications and commented on the manuscript at all stages.

## Competing interests

The authors declare no competing interests.

## Supplementary information

Supplementary information is available for this paper at XXX.